\documentclass{sig-alternate-10pt}
\usepackage[margin=1in]{geometry}
\usepackage{multirow}
\begin{document}
\pagestyle{empty}
\title{Business Intelligence for Small\\ and Middle-Sized Entreprises}
\numberofauthors{4} 
\author{
\alignauthor
Oksana Grabova\titlenote{This author also works at the Kharkiv National University of Economics, 9-a, pr.Lenina, 61001 Kharkov, Ukraine}\\
       \affaddr{University of Lyon (ERIC Lyon2)}\\
       \affaddr{5 av P. Mendes-France}\\
       \affaddr{69676 Bron Cedex, France}\\
       \email{oksana.grabova@eric.univ-lyon2.fr}
\alignauthor
Jerome Darmont\\
			 \affaddr{University of Lyon (ERIC Lyon2)}\\
       \affaddr{5 av P. Mendes-France}\\
       \affaddr{69676 Bron Cedex, France}\\
       \email{jerome.darmont@univ-lyon2.fr}
\alignauthor
Jean-Hugues Chauchat\\
			 \affaddr{University of Lyon (ERIC Lyon2)}\\
       \affaddr{5 av P. Mendes-France}\\
       \affaddr{69676 Bron Cedex, France}\\
       \email{jean-hugues.chauchat@univ-lyon2.fr}
       \and 
 \alignauthor Iryna Zolotaryova\\
			 \affaddr Kharkiv National University of Economics\\ 
       \affaddr{9-a, pr. Lenina}\\
       \affaddr{61001 Kharkov, Ukraine}\\
       \email{is@knue.edu.ua}
       }
\maketitle
\begin {abstract}
Data warehouses are the core of decision support systems, which nowadays are used by all kind of enterprises in the entire world. Although many studies have been conducted on the need of decision support systems (DSSs) for small businesses, most of them adopt existing solutions and approaches, which are appropriate for large-scaled enterprises, but are inadequate for small and middle-sized enterprises. \\ 
Small enterprises require cheap, lightweight architectures and tools (hardware and software) providing online data analysis. In order to ensure these features, we review web-based business intelligence approaches. For real-time analysis, the traditional OLAP architecture is cumbersome and storage-costly; therefore, we also review in-memory processing. \\ 
Consequently, this paper discusses the existing approa- ches and tools working in main memory and/or with web interfaces (including freeware tools), relevant for small and middle-sized enterprises in decision making.
\end{abstract}

\section {Introduction}
During the last decade, data warehouses (DWs) have become an essential component of modern decision support systems in most companies of the world. In order to be competitive, even small and middle-sized enterprises (SMEs) now collect large volumes of information and are interested in business intelligence (BI) systems \cite{24}. SMEs are regarded as significantly important on a local, national or even global basis and they play an important part in the any national economy \cite{31}.
In spite of multiples advantages, existing DSSs frequently remain inaccessible or insufficient for SMEs because of the following factors: 
\begin{itemize}
	\item high price;
	\item high requirements for a hardware infrastructure;
  \item	complexity for most users;
  \item irrelevant functionality;
  \item low flexibility to deal with a fast changing dynamic business environment \cite{47};
  \item low attention to difference in data access necessity in SMEs and large-scaled enterprises.
\end{itemize}

In addition, many projects fail due to the complexity of the development process. Moreover, as the work philosophies of small and large-scaled enterprises are considerably different, it is not advisable to use tools destined to large-scaled enterprises. In short, "one size does not fit all" \cite{62}. Furthermore, there are a lot of problems in the identification of information needs of potential users in the process of building a data warehouse \cite{8}. \\
Thereby, SMEs require lightweight, cheap, flexible, simple and efficient solutions. To aim at these features, we can take advantage of light clients with web interfaces. For instance, web technologies are utilized for data warehousing by large corporations, but there is an even greater demand of such kind of systems among small and middle-sized enterprises. Usage of web technologies provides cheap software, because it eliminates the necessity for numerous dispersed applications, the necessity of deployment and maintenance of corporate network, and reduces training time. It is simple for end-users to utilize web-based solutions. In addition, a web-based architecture requires only lightweight software clients (i.e., web browsers).\\
Besides, there is a need for real-time data analysis, which induces memory and storage issues. Traditional OLAP (On line Analytical Processing) tools are often based on a cumbersome hardware and software architecture, so they require significant resources to provide a high performance. Their flexibility is limited by data aggregation. At the same time, in-memory databases provide significant performance improvements. Absence of disk I/O operations permits fast query response times. In-memo- ry databases do not require indexes, recalculation and pre-aggregations, thus system becomes more flexible because analysis is possible to a detailed level without its pre-definition. Moreover, according to analyst firms, "by 2012, 70\% of Global 1000 organizations will load detailed data into memory as the primary method to optimize BI application performance" \cite{56}.\\
Thus, our objective is to propose an original and adapted BI solutions for SMEs. To this aim, in a first step, we review in this paper the existing research related to this issue.\\
The remainder of this paper is organized as follows. In section 2, we first present and discuss web-based BI approaches, namely web data warehouses and web-based open source software for data warehousing. In Section 3, we review in-memory BI solutions (MOLAP, vector database-based BI software) and technologies that can support it (in-memory and vector databases). We finally conclude this paper in Section 4 and provide our view on how the research and technologies surveyed in this paper can be enhanced to fit SME's BI needs.
\section {Web-powered BI}
The Web has become the platform of choice for the delivery of business applications for large-scaled entreprises as well as for SMEs. Web warehousing is a recent approach that merges data warehousing and business intelligence systems with web technologies \cite{44}. In this section, we present and discuss web data warehousing approaches, their features, advantages and possibilities, as well as their necessity and potential for SMEs.
\subsection {Web warehousing}
\subsubsection{General information}
There are two basic definitions of web warehousing. The first one simply states that web warehouses use data from the Web. The second concentrates on the use of web technologies in data warehousing. We focus on second definition in our paper.\\
Web-data warehouses inherit a lot of characteristics from traditional data warehouses, including: data are organized around major subjects in the enterprise; information is aggregated and validated; data is represented by times series, not by current status.\\
Web-based data warehouses nonetheless differ from traditional DWs. Web warehouses organize and manage the stored items, but do not collect them \cite{44}. Web-based DW technology changes the pattern of users accessing to the DW: instead of accessing through a LAN (Local Area Network), users access via Internet/Intranet \cite{27}.\\
Specific issues raised by web-based DW include unrealistic user expectations, especially in terms of how much information they want to be able to access from the Web; security issues; technical implementation problems related to peak demand and load problems \cite{30}.\\
Eventually, web technologies make data warehouses and decision support systems friendlier to users. They are often used in data warehouses only to visualize information \cite{53}. At the same time, web technology opens up multiple information formats, such as structured data, semi-structured data and unstructured data, to end-users. This gives a lot of possibilities to users, but also creates a problem known as data heterogeneity management \cite{6}. \\
Another important issue is the necessity to view the Web as an enormous source of business data, without whose enterprises loose a lot of possibilities. Owing to the Web, business analysts can access large external to enterprise information and then study competitor's movements by analyzing their web site content, can analyze customer preferences or emerging trends \cite{70}. So, e-business technologies are expected to allow SMEs to gain capabilities that were once the preserve of their larger competitors \cite{31}. However, most of the information in the Web is unstructured, heterogeneous and hence difficult to analyze \cite{24}.\\
Among web-technologies used in data warehousing, we can single out web browsers, web services and XML. Usage of web browser offers some advantages over traditional warehouse interface tools \cite{6,5}:
\begin{itemize}
	\item	cheapness and simplicity of web browser installation and use;
	\item	reduction of system training time;
	\item	elimination of problems posed by operating systems;
	\item	low cost of deployment and maintenance;
	\item	elimination of necessity for numerous dispersed applications;
	\item	possibility to open data warehouse to business partners over an extranet.
	\end{itemize}
Web warehouses can be divided into two classes: XML document warehouses and XML data warehouses. We present them in sections 2.1.2. and 2.1.3. respectively. We also introduce OLAP on XML data (XOLAP) in section 2.1.4. We finish this section by web-based paradigm known as cloud computing (section 2.1.5). Section 2.2. finally pre- sents web-based open source software for data warehousing analysis.
\subsubsection{XML document warehouses}
An XML document warehouse is a software framework for analyzing, sharing and reusing unstructured data (texts, multimedia documents, etc.). Unstructured data processing takes an important place in enterprise life because unstructured data are larger in volume than structured data, are more difficult to analyze, and are an enormous source of raw information.\\
Representing unstructured or semi-structured data with traditional data models is very difficult. For example, relational models such as star and snowflake schemas are semantically poor for unstructured data. Thus, Nassis et al. utilize object-oriented concepts to develop a conceptual model for XML document warehouses \cite{32}. They use UML diagrams to build hierarchical conceptual views. By combination of object oriented concepts and XML Schema, they build the xFACT repository. 
\subsubsection{XML data warehouses}
In contrast to XML document warehouses, XML data warehouses focus on structured data. XML data warehouse design is possible from XML sources \cite{3}. In this case, it is necessary to translate XML data into a relational schema by XML schema \cite{3,9}. 
Xyleme is one of the first projects aimed at XML data warehouse design \cite{48}. It collects and archives web XML documents into a dynamic XML warehouse.\\
Some more recent approaches are based on classical warehouse schemas. Pokorny adapts the traditional star schema with explicit dimension hierarchies for XML environments by using Document Type Definition (DTD) \cite{37}. Boussa\"{i}d et al. define data warehouse schemas via XML schema in a methodology named X-Warehousing \cite{9}. Golfarelli proposes a semi-automatic approach for building the conceptual schema for a data mart starting directly from XML sources \cite{16}. This work elaborates the concept of Dimensional Fact Model.
Baril and Bellahsene propose a View Model from XML Documents implemented in the DAWAX (Data Warehouse for XML) system \cite{2}.  View specification mechanism allows filtering data to be stored. N{\o}rv{\aa}g introduces a temporal XML data warehouses to query historical document versions and query changes between document versions \cite{33}. N{\o}rv{\aa}g et al. also propose TeXOR, a temporal XML database system built on top of an object-relational database system \cite{34}. Finally, Zhang et al. propose an approach, named X-Warehouse, to materialize data warehouses based on frequent query patterns represented by Frequent Pattern Trees \cite{45}.
\subsubsection{XOLAP}
Some recent research attempts to perform OLAP analysis over XML data. In order to support OLAP queries and to be able to construct complex analytic queries, some researches extend the XQuery language with aggregation features \cite{4}.\\
Wiwatwattana et al. also introduce an XQuery cube operator, X\^{ }3 \cite{50}, Hachicha et al. also propose a similar operator, but based on TAX (Tree Algebra for XML)\cite{19}. 
\subsubsection{Cloud computing}
Another, increasingly popular web-based solution is cloud computing. Cloud computing provides access to large amounts of data and computational resources through a variety of interfaces \cite{64}. It is provided as services via cloud (Internet). These services delivered through data centers are accessible anywhere. Besides, they allow the rise of cloud analytics \cite{65}.\\
The main consumers of cloud computing are small enterprises and startups that do not have a legacy of IT investments to manage \cite{67}. Cloud computing-based BI tools are rather cheap for small and middle-sized enterprises, because they provide no need of hardware and software maintenance \cite{68} and their prices increase according to required data storages. Contrariwise, cloud computing does not allow users to physically possess their data storage. It causes user dependence on the cloud computing provider, loss of data control and data security. In conclusion, most cloud computing-based BI tools do not fit enterprise requirements yet.
\subsubsection{Discussion}
Data storage and analysis interface solutions should be easily deployed in a small organization at low cost, and thus be based on web technologies such as XML and web services.
Web warehousing is rather recent, but a popular direction that provides a lot of advantages, especially in data integration. Web-based tools provide light interface. Thereby, their usage by small and middle-sized enterprises is limited. Existing cloud-based BI tools are appropriated for small and middle-sized enterprises with respect to price and flexibility. However, they are so far enterprise-friendly and are in need of data security enhancements.
\subsection{Web-based open source software}
In this section, we focus on ETL(Extraction Transformation Loading) tools, OLAP servers and OLAP clients. Their characteristics are summarized in Table 1.
\subsubsection{ETL}
\begin{table*}
	\centering
		\begin{tabular} {|l|p{1.5cm}|p{2.5cm}|p{1.5cm}|p{1.5cm}|p{6cm}|}
		\hline
		&&Tools&Platform&License&Particular features\\
		\hline
		\multirow{3}{*}{ETL}
			& \multirow{2}{*}{ROLAP} &Clover.ETL&Java&LGPL&does not have an open source GUI; uses its own TL language for data transformations\\ \cline{3-6}
		  		&&JasperETL&Java&GPL&generated code - Java or Perl; can use CRM systems as data sources\\ \cline{2-6}
		&MOLAP&Palo ETL Server&Java&GPL&does not have a GUI for a while; parallel jobs are not supported\\
		\hline
		\multirow{5}{*}{OLAP}
			& \multirow{2}{*}{servers}&Mondrian&Java&CPL&ROLAP-based; data cubes via XML\\ \cline{3-6}
		  		&&Palo&Java&GPL&MOLAP-based; works in memory; data cubes via Excel add-in\\ \cline{2-6}
		&\multirow{3}{*}{clients}&FreeAnalysis&Java&MPL&works with servers that use XMLA, e.g., Modrian\\ \cline{3-6}
			&&JPalo&Java&GPL&works with the Palo server\\ \cline{3-6}
			&&PocOLAP&Java&LGPL&\\ \cline{2-6}
		\hline
		\end{tabular}
	\caption{Web-based open source software}
	\label{tab:WebBasedOpenSourceSoftware}
\end{table*}
Web-based free ETL tools are in most cases ROLAP (Relational OLAP, discussed in Section 3.1.1.)-oriented. ROLAP-oriented ETL tools allow user to define and create data transformations in Java (JasperETL) or in TL (Clover.ETL)\footnote{http://www.cloveretl.com}. Singular MOLAP (Multidimensional OLAP, discussed in Section 3.1.1.)-oriented ETL Palo defines the ETL process either via web interfaces or via XML structures for experts. All studied ETL tools configure heterogeneous data sources and complex file formats. They interact with differents DBMSs (DataBase Management Systems). Some of the tools can also extract data from ERP (Enterprise Resource Planning) and CRM (Customer Relationship Management) systems \cite{57}. 
\subsubsection{OLAP}
In this section we review OLAP servers as well as OLAP clients. All studied OLAP severs use the MDX (Multi-Dimensional eXpression) language for  aggregating tables. They parse MDX into SQL to retrieve answers to dimensional queries. All reviewed OLAP servers exists for Java, but a Palo exists also for .NET, PHP, and C. Moreover, Palo is an in-memory Multidimensional OLAP database server\footnote{http://www.jedox.com/en/products/palo\_olap\_server}. Mondrian schemas are represented in XML files\footnote{http://mondrian.pentaho.org/}. Mondrian Pentaho Sever is used by different OLAP clients, e.g., FreeAnalysis.\\
All studied OLAP clients are Java applications. They usually run on client, but tools also exist that run on web servers\cite{57}. So far, only PocOLAP is a lightweight, open source OLAP solution. 
\subsubsection{Discussion}
The industrial use of open source business intelligence tools is becoming increasingly common, but it is still not as wide- spread as for other types of software \cite{57}. Moreover, freeware OLAP systems often propose simple web-based interfaces. In addition, there are some web-based open source BI tools that work in memory.\\
Nowadays, there are three complete solutions, including ETL and OLAP: Talend OpenStudio, Mondrian Pentaho and Pa- lo. \\Among ETL tools, only  Palo is MOLAP-oriented. Not all of these tools provide free graphical user interfaces. All three represented ETL tools support Java. They can be implemented on different platforms.\\
Free web-based OLAP servers are used by different OLAP clients. The most extended and widely used is Mondrian Pentaho Server due to its functionality. All studied OLAP clients are Java applications. Most of them can be used with XMLA(XML for Analysis)-enabled sources. But they have not enough documentation.\\
Generally, web-based studied tools provide sufficient functionality, but they remain cumbersome due to traditional OLAP usage.
\section {In-memory BI solutions}
In the late eighties, main memory databases were researched by numerous authors. Thereafter, it has rarely been discussed because of limits of technologies at this time, but nowadays it takes back an important place in database technologies. 
\subsection {MOLAP}
\subsubsection {OLAP and MOLAP}
Before studying existing MOLAP approaches, we review general OLAP principles and definitions. The OLAP concept was introduced in 1993 by Codd. OLAP is an approach to quickly answer multidimensional analytical queries \cite{13}. In OLAP, a \textit{dimension} is a sequence of analyzed parameter values. An important goal of multidimensional modeling is to use dimensions to provide as much context as possible for \textit{facts} \cite{54}. Combinations of dimension values define a cube's \textit{cell}. A \textit{cube} stores the result of different calculations and aggregations. 
There are three variants of OLAP: MOLAP, ROLAP, Hybrid OLAP (HOLAP). We compare these approaches in table 2.\\
\begin{table*}[t]
\centering	
	\caption{Comparison of OLAP technologies}
	\label{tab:ComparisonOfOLAPTechnologies}
	\begin {tabular} {|l|p{3.5cm}|p{3.5cm}|p{5.5cm}|}
	\hline
	&MOLAP&ROLAP&HOLAP\\
	\hline
	Data storage & Multidimensional database& Relational database & Uses MOLAP technology to store higher-level summary data, a ROLAP system to store detailed data \\ 
	\hline
	Results sets & Stores in a MOLAP cube & Stores no results sets & Stores results sets, but not all\\ 
	\hline
	Capacity & Requires singificant capacity & Requires the least storage capacity & \\ \cline{1-3} 
	Performance & The fastest performance & The slowest performance &\\ \cline{1-3}
Dimensions & Minimum number & Maximum number & \multirow{-4}{6cm}{Compromise between performance, capacity, and permutations of dimensions available to a user}\\ \cline{1-3}
	\hline
	Vulnerability & Provides poor storage utilization, especially when the data set is sparse & Database design
recommended by ER diagrams are inappropriate for decision support systems &\\
	\hline
	Advantages &Fast query performance; automated computation of higher level aggregates of the data; array model provides natural indexing &No limitation on data volume; leverage functionalities inherent in relational databases
&Fast access at all levels of aggregation; compact aggregate storage;	dynamically updated dimensions;	easy aggregate maintenance\\
	\hline
	Disadvantages &	Data redundancy; querying models with dimensions of high cardinality is difficult
&Slow performance&Complexity - a HOLAP server must support MOLAP and ROLAP engines, tools to combine storage engines and operations. Functionality overlap - between storage and optimization techniques in ROLAP and MOLAP engines.
\\
	\hline
	\end {tabular}
\end{table*}
With respect to ROLAP and HOLAP, MOLAP provides faster computation time and querying \cite{42} due to a storage of all required data in the OLAP server. Moreover, it provides more space-efficient storage \cite{36}. \\Since the purpose of MOLAP is to support decision making and management, data cubes must contain sufficient information to support decision making and reply to every user expectation. In this context, researches try to improve three main aspects: response time (by new aggregations algorithms \cite{26}, new operators \cite{43}), query personalization, data analysis visualization \cite{24}. 
\subsubsection {Storage methods}
Researchers interested in MOLAP focus a lot on storage techniques. In addition, most researches choose MOLAP as the most suitable among OLAP-techniques for storage \cite{29}, although MOLAP requires significant storage capacity. 
According to Kudryavcev, there are three basic types of storage methods: semantic, syntactical, approximate \cite{23}. Syntactical approaches transform only data storage schemas. Semantic storage techniques transform cube structures. Approximate storage techniques compress initial data. One semantic storage technique is Quotient Cube. It consists in a semantic compression by partitioning the set of cells of a cube into equivalent classes, while keeping the cube's roll-up and drill-down semantics and lattice structure \cite{58}. The main objective of such approximating storage technique such as Wavelets is range-sum query optimization \cite{59}. In the syntactical approach DWARF, a cube is compressed by deleting redundant information \cite{60}. Data are represented as graphs with keys and pointers in graphs nodes. Data redundancy decrease is provided by an addressing and data storage improvement.
\subsubsection{Schema evolution}
There are a lot of works that bring up the problem of schema evolution, because working only with the latest version hides the existence of information that may be critical for data analysis. It is possible to classify these studies into two groups: updating models (mapping data in the last version) and tracking history models (saving schema evolution).\\
Other types of approaches look at the possibility for users to choose which presentation they want for query reponses. For instance, Body et al. proposed a novel temporal multidimensional model for supporting evolutions on multidimensional structures by introducing a set of temporal modes of presentation for dimensions in a star schema \cite{7}.
\subsubsection {Discussion}
Multidimensional OLAP is appropriate for decision making. It offers a number of advantages, including automatic aggregation, visual querying, and good query performance due to the use of pre-aggregation \cite{35}.Besides, MOLAP may be a good solution for the situations in which small to medium-sized DBs are the norm and application software speed is critical \cite{63}, because loading all data to the multidimensional format does not require significant time nor disk space. Nevertheless, MOLAP systems have different problems due to the complexity, time-consuming and necessity of an expert for cube rebuilding. If the user wants to change dimensions, the whole deployment process need to be redone (datamart schema, ETL process, etc.) \cite{47}.  \\
However, the cost of MOLAP tools does not fit the needs of small and middle-sized enterprises. In addition, MOLAP-based systems may encounter significant scalability problems. Moreover, MOLAP requires a cumbersome architecture, i.e., important software and hardware needs, the necessity of significant changes in work process to generate substantial benefits \cite{69}, and a considerable deployment time.
\subsection {Main Memory Databases}
\subsubsection{General information}
Main Memory Databases (MMDBs) entirely reside in main memory \cite{15} and only use a disk subsystem for backup \cite{17}. The concept of managing an entire database in main memory has been researched for over twenty years, and the benefits of such approaches have been well-understood in certain domains, such as telecommunications, security trading, applications handling high traffic of data, e.g., routers; real-time applications. However, it is only recently, with decreasing memory prices and the availability of 64-bit operating systems, that the size restrictions on in-memory databases have been removed and in-memory data management has become available for many applications \cite{25, 51}. When the assumption of disk-residency is removed, complexity is dramatically reduced. The number of machine instructions drops, buffer pool management disappears, extra data copies are not needed, index pages shrink, and their structure is simplified. Design becomes simpler and more compact, and queries are executed faster \cite{51}. Consequently, usage of main memory databases become advantageous in many cases: for hot data (frequently access, low data volume), for cold data (scarce access, in the case of voluminous data), in application requiring a short access/response time.\\
A second wave of applications using MMDB is currently appearing, e.g., FastDB, Dali from AT\&T Bell lab, TimesTen from Oracle. These systems are widely used in many applications such as HP intellect web flat already, Cisco VoIP call Proxy, the telecom system of Alcatel and Ericsson and so on \cite{14}. The high demand of MMDBs is provoked by the necessity of high reliability, high real-time capacity, high quantity of information throughput \cite{21}.\\
MMDBs have some advantages, including short response time, good transaction throughputs. MMDBs also leverage the decreasing cost of main memory.   Contrariwise, MMDB size is limited by size of RAM (Random Access Memory). Moreover, since data in main memory can be directly accessed by the processor, MMDBs suffer from data vulnerability, i.e., risk of data loss because of unintended accident due to software errors\cite{15}, hardware failure or other hazards.
\subsubsection{MMDB issues}
Although in-memory technologies provide high performance, scalability and flexibility to BI tools, they are still some open issues. MMDBs work in memory, therefore the main problems and challenges are recovery, commit processing, access methods and storage.  \\
There is no doubt that backups of memory resident databa- ses must be maintained on other storage than main memory in order to insure data integrity. In order to protect against failures, it is necessary to have a backup copy and to keep a log of transaction activity \cite{15}. In addition, recovery processing is usually the only MMDB component that deals with disk I/O, so it must be designed carefully \cite{21}. Existing research works do not share a common view of this problem. Some authors propose to use a part of stable main memory to hold the log. It provides short response time, but it causes a problem when logs are large. So, it is used for the precommit transactions. Group commits (e.g., a casual commit protocol \cite{25}) allow accumulating several transactions in memory before flushing them to the log disk. Nowadays, commit processing is especially important in distributed database systems because it is slow due to the fact that disk logging takes place at several sites \cite{25}. \\
Several different approaches of data storage exist for MMDBs. Initially, there have been a lot of attempts to use database partitioning techniques developed earlier for other types of databases. Gruenwald and Eich divide existing techniques as following: horizontal partitioning, group partitioning, single vertical partitioning, group vertical partitioning, and mixed partitioning \cite{17}. Only horizontal and single vertical partitioning are suitable for MMDBs and, as a result of this study, single vertical partitioning was chosen as the most efficient \cite{12}. B-trees and hashing are identified also as appropriate storage techniques for MMDBs. Hashing is not as space efficient as a tree, so it is rarely used \cite{71}. Finally, most researches agree to choose T-trees (a balanced index tree data structure optimized for cases where both the index and the actual data are fully kept in memory) as the main storage technique \cite{14, 15, 40}. T-trees indeed require less memory space and fewer CPU cycles than B-trees, so indexes are more economical.\\
Above-mentioned issues are important for BI environment: data coherence is strategic, performance is fundamental for on-line operations like OLAP. Choices of right storage and recovering techniques are crucial as it can damage data security and data integrity.
\subsubsection{MMDB Systems}
In this section we give an overview of MMDB systems. We particularly focus our discussion on the most recent systems such as Dali, FastDB, Kdb, IBM Cognos TM1 and TimesTen. \\
Among studied systems, we can distinguish a storage manager (the Dali system \cite{21}) and complete main memory data- base systems (FastDB, Kdb, TM1, TimesTen). Interfaces can be based on zero-footprint Web (IBM Cognos TM1\footnote{www-01.ibm.com/software/data/cognos/products/tm1}), standard SQL (TimesTen)\cite{51} or C++ (FastDB)\cite{22}. Most MMDBS feature SQL or SQL-like query language (FastDB, TimesTen). Kdb system uses its own language "q" for programming and querying \cite{66}.\\
IBM Cognos 8 BI and TimesTen are aimed at decision making in large corporations. 
Main MMDB disadvantages are interprocess communication absence and high storage requirements (Dali system) \cite{10}, limitation of server memory (TimesTen), client-server architecture is unsupportable (FastDB).  
\subsubsection{Discussion}
The main benefit of using MMDBs is short access/reponse time and good transaction throughput. But MMDBs are hampered by data vulnerability and security problems. Memory is not persistent, which means data loss in case of failure on the server. Security problems come from unauthorized access to data aimed at data corruption or theft.\\ 
So far, MMDBs are mainly used in real-time applications, telecommunications, but not commonly used for decision making.\\
In spite of a considerable research on MMDBs, there are some unresolved issues such as data security and safety and data processing efficiency. 
\subsection {Vector Databases}
\subsubsection{General information}
A vector table is built by transforming a file in the following way: every record represents all column values in a vector. Vector databases (VDBs) do not require indexes nor any complex database structure. Differences between vector and relational databases are summarized in table 2. 
\begin {table*}[t]
\centering
\caption{Relational and vector database chatacteristics}
\begin {tabular} {|l|l|l|}
\hline
 Characteristic & 
 Relational DB & Vector DB\\
 \hline
 Access to data & Sequential & Parallel\\
  \hline
 Data integrity & Foreign Keys & Multi-dimensional\\
 \hline
 Data relations stored in & Keys & Vectors\\
 \hline
 Data reuse & Not available & Built-in\\ \hline
 Metadata & System tables &  None\\
 \hline
 Speed (high volume) & Slow & Fast\\
 \hline
 Uniqueness & User Constraints & Built-in\\
 \hline
\end {tabular}
\end {table*}
In order to access data, relational DBMSs provide only sequential scan by columns and by rows. VDBs provide fast data access. Besides, relational DBs store large volumes of repeatable data due to data nature. For example, in a table of students, French nationality can be repeated in a great number. Contrariwise, in VDBs, this data is present only once. It provides significant data compression.\\The main principles of vector databases are data associations and data access by pointers. Vector database implementations allow elimination of data redundancy, because any possible pice of data is written once and it does not repeat itself. Such metadata as keys in the relational data model loose their interest in VDBs, because data associations are provided by pointers. Hence, VDBs do not consume as much space as relational DBs.
\subsubsection{VDB-based BI}
The main principle of vector database is that instead of dimension associations with OLAP cube there are associations between data. These associations are defined during data load process by matching up table columns having the same name. Usage of vector databases differs from classical warehousing: there is no predefinition of what a dimension is. Any piece of data is available as dimension and any piece of data is available as measure. So, it is not necessary to reconstruct data schema in the case of dimension change. As vector databases work in memory, VDB-based BI are endowed with instant data access. However, entreprises frequently hesitate to use VDB-based BI because of noninteroperability with SQL tools. \\ 
One BI tool that uses vector database deployment is QlikView\footnote{www.qlikview.com}. QlikView provides integrated ETL. It removes the need to pre-aggregate data. It is possible to change analysis axes any moment at any level of query detailing. Despite QlikView capacities, it has some limitations and disadvantages such as lack of a unified metadata view and of predicting models (QlikView's statistical analysis features are less developed than the in other BI tools). There is no specialization in visualization: QlikView provides a clean interface to analysts but it lacks advanced visualization features to help them graphically wade through complicated data. One of the QlikViews's features is an ability to automatically connect tables. But this can create some problems. When there are fields, which represent the same thing in different tables and they do not have the same name, it is necessary to rename them to connect them. When there are fields in different tables that have the same name, but not the same content and sense, a senseless connection is created. So it is necessary to delete this connection and reanalyze all the fields with the same names in order to distinguish the ones with different sense. QlikView provides a possibility for end-users to use integrated ETL and to construct their data schema themselves, which often leads to unsatisfactory results.  
\subsubsection{Discussion}
Vector databases hold the same advantages as others in-memory databases and are only limited by memory size.\\ 
VDB-based BI is a relatively new direction, but it is rather popular due to fast performance, great analysis capacity, unlimited number of dimensions, tables and measures and implementation easiness.\\ However, among features proposed by QlikView, there are disputable ones: automatic table connection, possibility to create a data schema by end-user. These characteristics do not cover different situations due to data aggregation complexity when data come from different sources. Such data have different refinement levels, different field names, etc. Consequently, providing to end-users the possibility to create data schemas can provoke an inadequate data schema, table connections, data loss as well as false data presence in database. Moreover, VDB-based BI tools are often blackboxes, meaning that we do not know what happens inside. Such models also lack flexibility. 
\section {Conclusion} 
Nowadays, BI becomes an essential part of any enterprise, even an SME. This necessity is caused by the increasing data volume indispensable for decision making. Existing solutions and tools are mostly aimed at large-scaled enterprises; thereby they are inaccessible or insufficient for SMEs because of high price, redundant functionality, complexity, and high hardware and software requirements. SMEs require solutions with light architectures that, moreover, are cheap and do not require additional hardware and software.\\
This survey discusses the importance of data warehousing for SMEs, presents the main characteristics and examples of web-based data warehousing, MOLAP systems and MMDBs. All these approaches have important disadvantages to be chosen as a unique decision support system: cumbersome architecture and complexity in MOLAP, data vulnerability in MMDBs, non-transparency and providing too large powers for users in VDB-based systems, security issues in cloud computing systems.\\
In this context, our research objective is to design BI solutions that are suitable for SMEs and avoid the aforementioned disadvantages.\\
Our idea is to work toward a ROLAP system that operates in-memory, i.e., to add in OLAP operators on top of an SQL-based MMDB. This should simplify a lot the in-memory OLAP architecture with respect to MOLAP. Choosing an open source MMDB system (such as FastDB) and using well-known ETL, modeling and analysis processes should also help avoid the "black box issue" of VDBs. Finally, storing business data as close to the user as possible mitigates security issues with respect to cloud BI. Problems will still remain, though (e.g., data vulnerability and need for backup, the design of adapted, in-memory indexes for OLAP), but we are confident we can address them in our future research.
 \section{Acknowledgments}
The authors would like to thank the French Ambassy in Ukraine for supporting this joint research work of the Kharkiv National University of Economics (Ukraine) and  the University of Lyon 2 (France).
\bibliographystyle{abbrv}	
\bibliography{art1}

\begin{thebibliography}{10}

\bibitem{68}
D.~J. Abadi.
\newblock {Data Management in the Cloud: Limitations and Opportunities}.
\newblock {\em IEEE Data Engineering Bulletin}, 32(1):3--12, March 2009.

\bibitem{65}
M.~Armbrust, A.~Fox, R.~Griffith, A.~D. Joseph, R.~H. Katz, A.~Konwinski,
  G.~Lee, D.~A. Patterson, A.~Rabkin, I.~Stoica, and M.~Zaharia.
\newblock {Above the Clouds: A Berkeley View of Cloud Computing}.
\newblock Technical Report UCB/EECS-2009-28, EECS Department, University of
  California, Berkeley, February 2009.

\bibitem{3}
M.~Banek, Z.~Skocir, and B.~Vrdoljak.
\newblock {{ Logical Design of Data Warehouses from XML }}.
\newblock In {\em ConTEL '05: Proceedings of the 8th international conference
  on Telecommunications}, volume~1, pages 289--295, 2005.

\bibitem{2}
X.~Baril and Z.~Bellahsene.
\newblock {Designinig and Managing an XML Warehouse}.
\newblock In {\em {XML Data Management: Native XML and XML-Enabled Database
  Systems}}, chapter~16, pages 455--473. {Addison-Wesley Professional}, 2003.

\bibitem{4}
K.~Beyer, D.~Chamberlin, L.~S. Colby, F.~\"{O}zcan, H.~Pirahesh, and Y.~Xu.
\newblock Extending {XQ}uery for analytics.
\newblock In {\em SIGMOD '05: Proceedings of the 2005 ACM SIGMOD international
  conference on Management of data}, pages 503--514, New York, NY, USA, 2005.
  ACM.

\bibitem{7}
M.~Body, M.~Miquel, Y.~B\'{e}dard, and A.~Tchounikine.
\newblock {A multidimensional and multiversion structure for OLAP
  applications}.
\newblock In {\em DOLAP '02: Proceedings of the 5th ACM international workshop
  on Data Warehousing and OLAP}, pages 1--6, New York, NY, USA, 2002. ACM.

\bibitem{8}
M.~B\"{o}hnlein and A.~U. vom Ende.
\newblock {\em {Business Process Oriented Development of Data Warehouse
  Structures}}, pages 3--21.
\newblock Physica: Heidelberg 2000, 2000.

\bibitem{9}
O.~Boussa\"{i}d, R.~BenMessaoud, R.~Choquet, and S.~Anthoard.
\newblock {X-Warehousing: an XML-Based Approach for Warehousing Complex Data}.
\newblock In {\em ADBIS '06: Proceedings of the 10th East-European Conference
  on Advances in Databases and Information Systems}, volume 4152 of {\em
  Lecture Notes in Computer Science}, pages 39--54, Heidelberg, Germany,
  September 2006. Springer.

\bibitem{10}
P.~Burte, B.~Aleman-meza, D.~B. Weatherly, R.~Wu, S.~Professor, and J.~A.
  Miller.
\newblock {Transaction Management for a Main-Memory Database}.
\newblock {\em The 38th Annual Southeastern ACM Conference, Athens, Georgia},
  pages 263--268, January 2001.

\bibitem{12}
Y.~C. Cheng, L.~Gruenwald, G.~Ingels, and M.~T. Thakkar.
\newblock {Evaluating Partitioning Techniques for Main Memory Database:
  Horizontal and Single Vertical}.
\newblock In {\em ICCI '93: Proceedings of the 5th International Conference on
  Computing and Information}, pages 570--574, Washington, DC, USA, 1993. IEEE
  Computer Society.

\bibitem{70}
W.~Chung and H.~Chen.
\newblock {Web-Based Business Intelligence Systems: A Review and Case Studies}.
\newblock In G.~Adomavicius and A.~Gupta, editors, {\em {Business Computing}},
  volume~3, chapter~14, pages 373--396. {Emerald Group Publishing}, 2009.

\bibitem{14}
Y.~Cui and D.~Pi.
\newblock {SQLmmdb: An Embedded Main Memory Database Management System}.
\newblock {\em Information Technology Journal}, 6(6):872--878, 2007.

\bibitem{13}
S.~C. E.F.~Codd and C.~Salley.
\newblock {Providing OLAP to User-Analysts: An IT Mandate}, 1993.

\bibitem{15}
H.~Garcia-Molina and K.~Salem.
\newblock Main memory database systems: An overview.
\newblock {\em IEEE Transactions on Knowledge and Data Engineering},
  4:509--516, 1992.

\bibitem{16}
M.~Golfarelli, S.~Rizzi, and B.~Vrdoljak.
\newblock {Data warehouse design from XML sources}.
\newblock In {\em DOLAP '01: Proceedings of the 4th ACM international workshop
  on Data warehousing and OLAP}, pages 40--47, New York, NY, USA, 2001. ACM.

\bibitem{17}
L.~Gruenwald and M.~H. Eich.
\newblock Choosing the best storage technique for a main memory database
  system.
\newblock In {\em JCIT '90: Proceedings of the 5th Jerusalem conference on
  Information technology}, pages 1--10, Los Alamitos, CA, USA, 1990. IEEE
  Computer Society Press.

\bibitem{19}
M.~Hachicha, H.~Mahboubi, and J.~Darmont.
\newblock {Expressing OLAP operators with the TAX XML algebra}.
\newblock In {\em DataX-EDBT '08: 3rd International Workshop on Database
  Technologies for Handling XML Information on the Web}, pages 61--66, March
  2008.

\bibitem{53}
H.~A.~A. Hafez and S.~Kamel.
\newblock {Web-Based Data Warehouse in the Egyptian Cabinet Information and
  Decision Support Center}.
\newblock In R.~Meredith, G.~Shanks, D.~Arnott, and S.~Carlsson, editors, {\em
  DSS'04: The IFIP International Conference on Decision Support in an Uncertain
  and Complex World}, pages 402--409. Monash University, Australia (CD Rom),
  July 2004.

\bibitem{6}
C.~Hsieh and B.~Lin.
\newblock Web-based data warehousing: current status and perspective.
\newblock {\em The Journal of Computer Information Systems}, 43:1--8, January
  2002.

\bibitem{21}
H.~V. Jagadish, D.~F. Lieuwen, R.~Rastogi, A.~Silberschatz, and S.~Sudarshan.
\newblock {Dal\'{\i}: A High Performance Main Memory Storage Manager}.
\newblock In {\em VLDB '94: Proceedings of the 20th International Conference on
  Very Large Data Bases}, pages 48--59, San Francisco, CA, USA, 1994. Morgan
  Kaufmann Publishers Inc.

\bibitem{54}
R.~Kimball and M.~Ross.
\newblock {\em {The Data Warehouse Toolkit: the complet guide to dimensional
  modeling}}.
\newblock Wiley Computer Publishing, 2002.

\bibitem{22}
K.~Knizhnik.
\newblock {FastDB Main Memory Database Management System}.
\newblock Technical report, Research Computer Center of Moscow State
  University, Russia, March 1999.

\bibitem{23}
Y.~Kudryavcev.
\newblock Efficient algorithms for {MOLAP} data storage and query processing.
\newblock In {\em {SYRCoDIS '06: Proceedings of the 3rd Spring Colloquium for
  Young Researchers in Databases and Information Sytems}}, page~5, Moscow,
  Russia, 2006.

\bibitem{66}
{Kx Systems}.
\newblock {The kdb+ Database White Paper. A unified database for streaming and
  historical data}, 2009.
\newblock Retrieved September 1, 2010 from http://kx.com/papers.

\bibitem{58}
L.~V.~S. Lakshmanan, J.~Pei, and J.~Han.
\newblock Quotient cube: how to summarize the semantics of a data cube.
\newblock In {\em VLDB '02: Proceedings of the 28th international conference on
  Very Large Data Bases}, pages 778--789. VLDB Endowment, 2002.

\bibitem{24}
G.~Lawton.
\newblock {Users Take a Close Look at Visual Analytics}.
\newblock {\em Computer}, 42(2):19--22, 2009.

\bibitem{25}
I.~Lee, H.~Y. Yeom, and T.~Park.
\newblock {A New Approach for Distributed {M}ain {M}emory {D}atabase {S}ystems:
  A Causal Commit Protocol}.
\newblock {\em IEICE Transactions on Information and Systems},
  E87-D(1):196--204, January 2004.

\bibitem{26}
Y.-K. Lee, K.-Y. Whang, Y.-S. Moon, and I.-Y. Song.
\newblock An aggregation algorithm using a multidimensional file in
  multidimensional {OLAP}.
\newblock {\em Information Sciences}, 152(1):121--138, June 2003.

\bibitem{59}
D.~Lemire.
\newblock {Wavelet-Based Relative Prefix Sum Methods for Range Sum Queries in
  Data Cubes}.
\newblock In {\em CASCON '02: Proceedings of the 2002 conference of the Centre
  for Advanced Studies on Collaborative research}, page~6. IBM Press, October
  2002.

\bibitem{27}
Z.~Luo, Z.~Kaisong, X.~Hongxia, and Z.~Kaipeng.
\newblock {The Data Warehouse Model Based on Web Service Technology}.
\newblock {\em Journal of Communication and Computer}, 2(1):26--31, January
  2005.

\bibitem{29}
E.~Malinowski and E.~Zim\'{a}nyi.
\newblock Hierarchies in a multidimensional model: from conceptual modeling to
  logical representation.
\newblock {\em Data \& Knowledge Engineering}, 59(2):348--377, 2006.

\bibitem{69}
M.~McDonald.
\newblock Light weight vs heavy weight technologies, the difference matters.
\newblock Gartner, March 2010.

\bibitem{5}
A.~Mehedintu, I.~Buligiu, and C.~Pirvu.
\newblock Web-enabled {D}ata {W}arehouse and {D}ata {W}ebhouse.
\newblock {\em Revista Informatica Economica}, 1(45):96--102, 2008.

\bibitem{31}
R.~Mullins, Y.~Duan, D.~Hamblin, P.~Burrell, H.~Jin, G.~Jerzy, Z.~Ewa, and
  B.~Aleksander.
\newblock {A Web Based Intelligent Training System for SMEs}.
\newblock {\em The Electronic Journal of e-Learning}, 5:39--48, 2007.

\bibitem{32}
V.~Nassis, W.~Rahayu, R.~Rajugan, and T.~Dillon.
\newblock Conceptual design of {XML} document warehouses.
\newblock In {\em DaWak '04: Proceedings of the 6th International on Data
  Warehousing and Knowledge Discovery}, pages 1--14, 2004.

\bibitem{33}
K.~N{\o}rv{\aa}g.
\newblock {Temporal XML Data Warehouses: Challenges and Solutions}.
\newblock In {\em Proceedings of Workshop on Knowledge Foraging for Dynamic
  Networking of Communities and Economies(in conjunction with
  EurAsia-ICT'2002)}, October 2002.

\bibitem{34}
K.~N{\o}rv{\aa}g, M.~Limstrand, and L.~Myklebust.
\newblock {TeXOR: Temporal XML Database on an Object-Relational Database
  System}.
\newblock In M.~Broy and A.~V. Zamulin, editors, {\em Ershov Memorial
  Conference}, volume 2890 of {\em Lecture Notes in Computer Science}, pages
  520--530. Springer, 2003.

\bibitem{64}
D.~Nurmi, R.~Wolski, C.~Grzegorczyk, G.~Obertelli, S.~Soman, L.~Youseff, and
  D.~Zagorodnov.
\newblock {The Eucalyptus Open-Source Cloud-Computing System}.
\newblock In {\em CCGRID '09: Proceedings of the 2009 9th IEEE/ACM
  International Symposium on Cluster Computing and the Grid}, pages 124--131,
  Washington, DC, USA, 2009. IEEE Computer Society.

\bibitem{35}
D.~Pedersen, K.~Riis, and T.~B. Pedersen.
\newblock {XML-Extended OLAP Querying}.
\newblock In {\em SSDBM '02: Proceedings of the 14th International Conference
  on Scientific and Statistical Database Management}, pages 195--206,
  Washington, DC, USA, 2002. IEEE Computer Society.

\bibitem{36}
T.~B. Pedersen and C.~S. Jensen.
\newblock {Multidimensional Database Technology}.
\newblock {\em Computer}, 34(12):40--46, 2001.

\bibitem{37}
J.~Pokorn\'{y}.
\newblock {XML Data Warehouse: Modelling and Querying}.
\newblock In {\em Baltic DB\&IS '02: Proceedings of the 5th International
  Baltic Conference on Databases and Information Systems}, pages 267--280.
  Institute of Cybernetics at Tallin Technical University, 2002.

\bibitem{30}
D.~J. Power and S.~Kaparthi.
\newblock {Building Web-based Decision Support Systems}.
\newblock {\em Studies in Informatics and Control}, 11(4):291--302, December
  2002.

\bibitem{71}
J.~Rao and K.~A. Ross.
\newblock {Cache Conscious Indexing for Decision-Support in Main Memory}.
\newblock In {\em {VLDB '99: Proceedings of the 25th International Conference
  on Very Large Data Bases}}, pages 78--89, San Francisco, CA, USA, 1999.
  Morgan Kaufmann Publishers Inc.

\bibitem{40}
R.~Rastogi, S.~Seshadri, P.~Bohannon, D.~W. Leinbaugh, A.~Silberschatz, and
  S.~Sudarshan.
\newblock {Logical and Physical Versioning in Main Memory Databases}.
\newblock In {\em VLDB '97: Proceedings of the 23rd International Conference on
  Very Large Data Bases}, pages 86--95, San Francisco, CA, USA, 1997. Morgan
  Kaufmann Publishers Inc.

\bibitem{63}
P.~Rob and C.~Coronel.
\newblock {\em Database systems: design, implementation, and management}.
\newblock Cengage Learning, 2007.

\bibitem{43}
G.~Sathe and S.~Sarawagi.
\newblock {Intelligent Rollups in Multidimensional OLAP Data}.
\newblock In {\em VLDB '01: Proceedings of the 27th International Conference on
  Very Large Data Bases}, pages 531--540, San Francisco, CA, USA, 2001. Morgan
  Kaufmann Publishers Inc.

\bibitem{56}
K.~Schlegel, M.~A. Beyer, A.~Bitterer, and B.~Hostmann.
\newblock {BI Applications Benefit From In-Memory Technology Improvements}.
\newblock Gartner, October 2006.

\bibitem{42}
F.~Silvers.
\newblock {\em {Building and Maintaining a Data Warehouse}}.
\newblock Auerbach Publications, 2008.

\bibitem{60}
Y.~Sismanis, A.~Deligiannakis, N.~Roussopoulos, and Y.~Kotidis.
\newblock {DWARF: shrinking the PetaCube}.
\newblock In {\em SIGMOD '02: Proceedings of the 2002 ACM SIGMOD international
  conference on Management of data}, pages 464--475, New York, NY, USA, 2002.
  ACM.

\bibitem{67}
J.~Staten.
\newblock Is cloud computing ready for the enterprise?
\newblock {Forrester Research}, March 2008.
\newblock Retrieved September 1, 2010 from
  http://www.forrester.com/rb/Research/is\_cloud
  \_computing\_ready\_for\_enterprise/q/id/44229/t/2.

\bibitem{62}
M.~Stonebracker and N.~Hachem.
\newblock {The End of an Architectual Era (It's Time for a Complete Rewrite)}.
\newblock In {\em VLDB'07: Proceedings of the 33rd international conference on
  Very Large Data Bases}, pages 1150--1160, 2007.

\bibitem{44}
X.~Tan, D.~C. Yen, and X.~Fang.
\newblock {Web warehousing: Web technology meets data warehousing}.
\newblock {\em Technology in Society}, 25:131--148, January 2003.

\bibitem{57}
C.~Thomsen and T.~B. Pedersen.
\newblock {A Survey of Open Source Tools for Business Intelligence}.
\newblock {\em International Journal of Data Warehousing and Mining},
  5(3):56--75, jul-sep 2009.

\bibitem{51}
C.~TimesTen~Team.
\newblock In-memory data management for consumer transactions the timesten
  approach.
\newblock {\em SIGMOD Record}, 28(2):528--529, 1999.

\bibitem{50}
N.~Wiwatwattana, H.~V. Jagadish, L.~V.~S. Lakshmanan, and D.~Srivastava.
\newblock {X\^{ }3: A Cube Operator for XML OLAP}.
\newblock {\em IEEE 23rd International Conference on Data Engineering}, pages
  916--925, 2007.

\bibitem{47}
G.~Xie, Y.~Yang, S.~Liu, Z.~Qiu, Y.~Pan, and X.~Zhou.
\newblock {EIAW: Towards a Business-friendly Data Warehouse Using Semantic Web
  Technologies}.
\newblock In K.~Aberer, K.-S. Choi, N.~Noy, D.~Allemang, K.-I. Lee, L.~J.~B.
  Nixon, J.~Golbeck, P.~Mika, D.~Maynard, G.~Schreiber, and P.~Cudre-Mauroux,
  editors, {\em ISWC/ASWC '07: Proceedings of the 6th International Semantic
  Web Conference and 2nd Asian Semantic Web Conference}, volume 4825 of {\em
  LNCS}, pages 851--904, Berlin, Heidelberg, November 2007. Springer Verlag.

\bibitem{48}
L.~Xyleme.
\newblock {A dynamic warehouse for XML data of the Web}.
\newblock {\em IEEE Data Engineering Bulletin}, 24:40--47, 2001.

\bibitem{45}
J.~Zhang, W.~Wang, H.~Liu, and S.~Zhang.
\newblock X-warehouse: building query pattern-driven data.
\newblock In {\em WWW '05: Special interest tracks and posters of the 14th
  international conference on World Wide Web}, pages 896--897, New York, NY,
  USA, 2005. ACM.

\end{thebibliography}
\end{document}